\documentclass[conference]{IEEEtran}
\IEEEoverridecommandlockouts
\usepackage{cite}
\usepackage{amsmath,amssymb,amsfonts}
\usepackage{algorithmic}
\usepackage{graphicx}
\usepackage{textcomp}
\usepackage{subcaption}
\usepackage{xcolor}
\def\BibTeX{{\rm B\kern-.05em{\sc i\kern-.025em b}\kern-.08em
    T\kern-.1667em\lower.7ex\hbox{E}\kern-.125emX}}
\begin{document}

\title{Impact of Medium and Heavy-Duty Electric Vehicle Electrification on Distribution System Stability\\

}

\author{
\IEEEauthorblockN{1\textsuperscript{st} Ali Hassan}
\IEEEauthorblockA{\textit{Department of Electrical and Computer Engineering} \\
\textit{University of Michigan – Dearborn}\\
Dearborn, MI, USA \\
alihssn@umich.edu}
\and
\IEEEauthorblockN{2\textsuperscript{nd} Wanshi Hong}
\IEEEauthorblockA{\textit{Lawrence Berkeley National Laboratory}\\
Berkeley, CA, USA \\
wanshihong@lbl.gov}
\and
\IEEEauthorblockN{3\textsuperscript{rd} Bin Wang}
\IEEEauthorblockA{\textit{Lawrence Berkeley National Laboratory}\\
Berkeley, CA, USA \\
wangbin@lbl.gov}
\and
\IEEEauthorblockN{4\textsuperscript{th} Wencong Su}
\IEEEauthorblockA{\textit{Department of Electrical and Computer Engineering} \\
\textit{University of Michigan – Dearborn}\\
Dearborn, MI, USA \\
wencong@umich.edu}
}
\maketitle

\begin{abstract}
Medium and heavy-duty (MHD) commercial vehicles contribute significantly to carbon emissions, accounting for 21\%  of the total emissions in the transportation sector. To curb this, U.S. government is increasingly focusing on
achieving 100\% fleet electrification over the next decade. However, the integration of megawatt-scale charging stations designed for MHD vehicles poses challenges to the stability of secondary distribution systems. This study investigates the impact of megawatt-scale charging station loads
on a benchmark IEEE 33-bus distribution system using real data from the HEVI-LOAD software for MHD electrification planning developed by Lawrence Berkeley National Laboratory (LBNL). The results reveal significant violations of per-unit (p.u.) voltage values at various nodes of the distribution system, indicating that substantial upgrades to the distribution infrastructure will be necessary to accommodate the projected MHDEV charging loads and meet electrification targets.
\end{abstract}


\section{Introduction}
Around the world, governments have adopted aggressive vehicle electrification policies to curb climate change and mitigate greenhouse gas (GHG) emissions. In the United States, heavy-duty commercial vehicles account for 21\% of GHG emissions in the transportation sector\cite{b1}. In California, the Advanced Clean Truck Regulation (ACT) requires manufacturers to sell increasing percentages of zero-emission medium- and heavy-duty (MHD) vehicles from 2024 to 2035 \cite{b2}. The US National Blueprint for Transportation Decarbonization by the Department of Energy contains a national plan to achieve 30\%  zero-emission medium and heavy-duty electric vehicle (MHDEV) sales by 2030 and 100\% sales by 2040 \cite{b3}. Average battery capacities in commercial trucks can vary from 330 kWh to 660 kWh \cite{b4}, significantly larger due to heavy payload requirements. Commercial vehicles belonging to short-haul delivery fleets can be slowly charged at the charging stations in the depot while parked overnight. However, the long-haul vehicles need to be charged along the highway corridors, and due to limited driver rest times, these charging stations are required to fast-charge the MHD vehicles. Such charging stations are categorized as Mega-Watt (MW) scale charging stations or high-power charging stations (HPCSs). The high power requirements of HPCS create severe stress on the power distribution system.

The driving patterns and charging behaviors of low-duty electric vehicles (LDEVs) are significantly different from those of MHDEVs. The impacts of light-duty electric vehicles (LDEVs) on the electric grid have been thoroughly researched, but the impacts of MHDEVs on the distribution grid is less explored. In \cite{b5}, the authors explored the impacts of depot charging stations on 36 real-world distribution systems and found that most distribution systems can accommodate slow-charging MHDEV loads. Optimization of MHDEV charging schedules for fleet MHDEVs can decrease grid stress \cite{b6},\cite{b7}. A study found that 11\% of MHDEVs charging simultaneously can cause significant voltage violations in the Texas transmission system, resulting in severe reliability problems for the grid\cite{b8}. There is a clear gap in understanding the impact of MHDEVs on the secondary distribution system. This study is important because it will aid in the planning and the requirements of distribution system degradation that can take years and cost a lot to the utilities.

This paper explores the impacts of MHDEV load on the secondary distribution grid. The main contributions of this paper are:
\begin{enumerate}
\item Real-world data from the HEVI-LOAD software\cite{b8} is combined with a queue model $M/M/c$ to build a realistic load demand profile of MHDEVs.
\item The IEEE 33-bus distribution system is modified to reflect real residential loads.
\item Power flow simulations are performed on the IEEE 33 bus system to assess the voltage stability on various nodes under the presence of MW-scale charging load.
\end{enumerate}

\section{Methodology}

\subsection{Real world MHD charging data}
\begin{figure}
    \centering
    \includegraphics[width=1\linewidth]{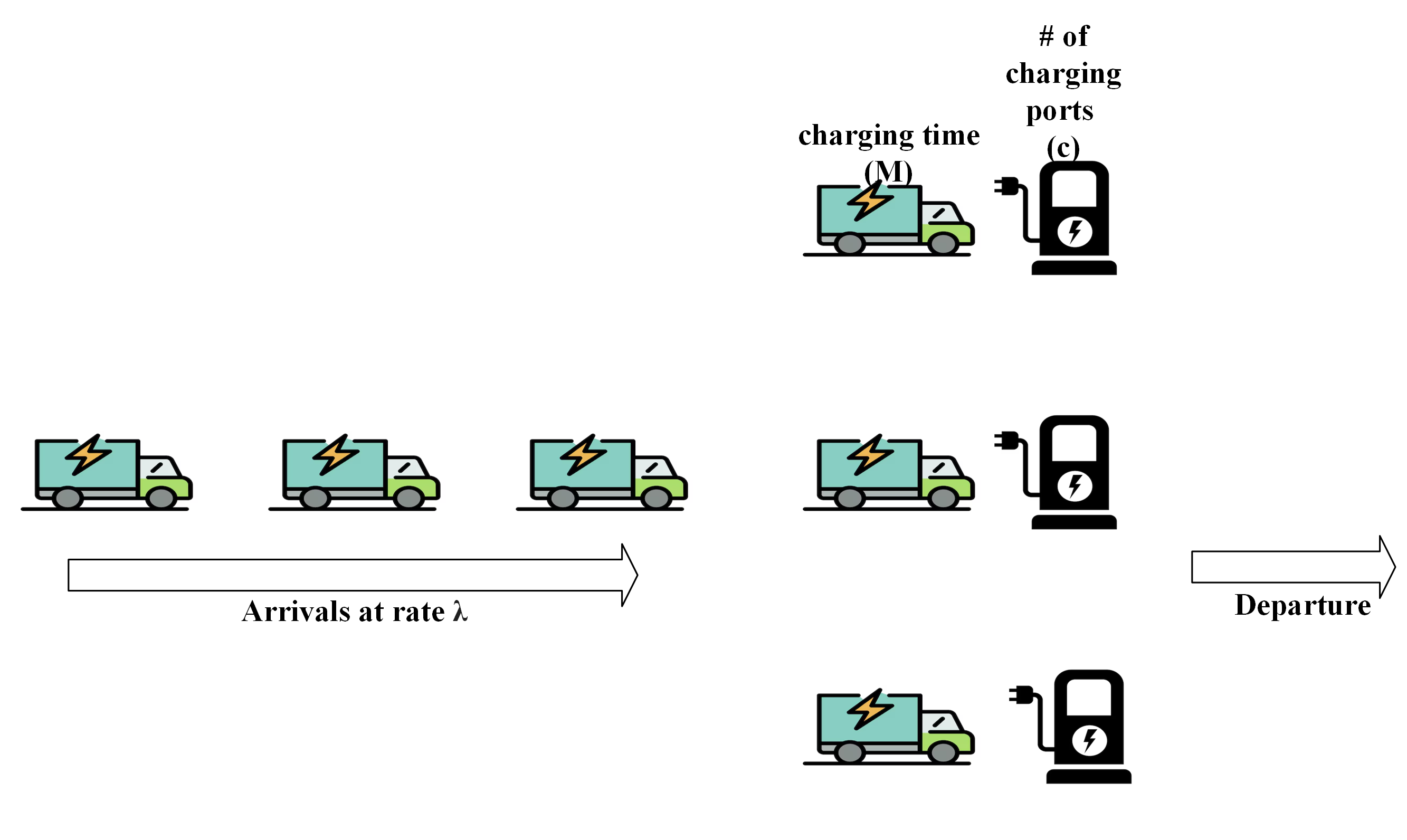}
    \caption{M/M/c queue model at HPC station}
    \label{fig:mmc}
\end{figure}

\begin{figure}[!t]
    \centering
    \begin{subfigure}[b]{1\columnwidth}
        \centering
        \includegraphics[width=\textwidth]{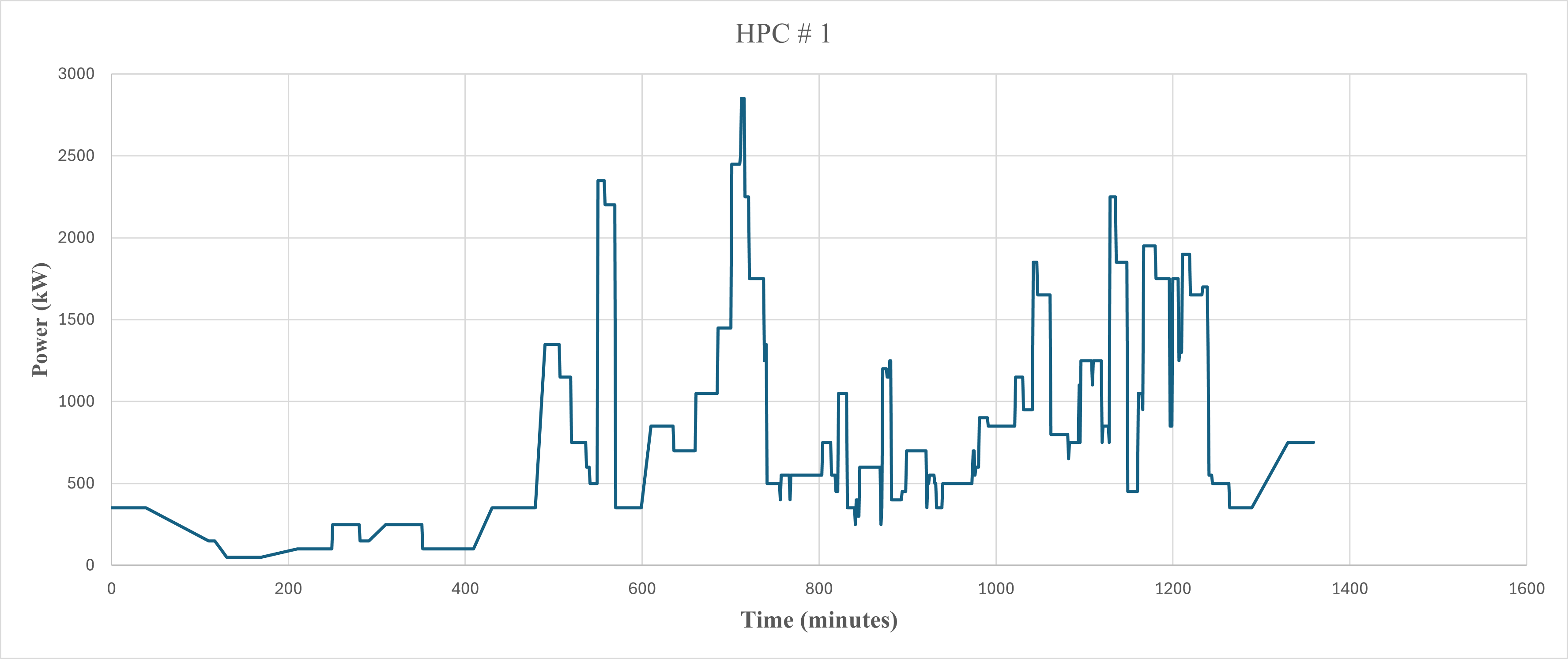}
        \caption{Load profile for HPCS 1}
        \label{fig:subfig1}
    \end{subfigure}
    
    \vspace{0.3cm} 
    
    \begin{subfigure}[b]{1\columnwidth}
        \centering
        \includegraphics[width=\textwidth]{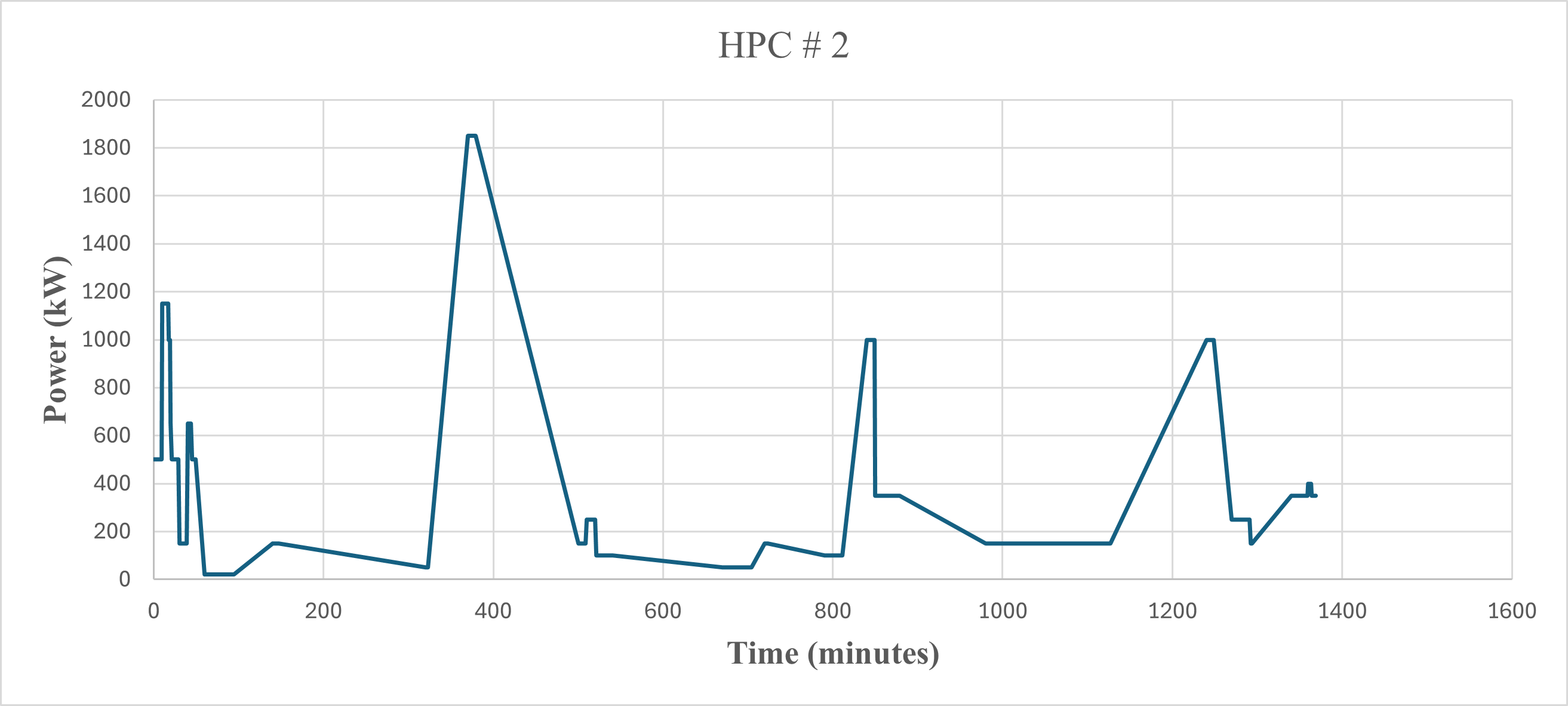}
        \caption{Load profile for HPCS 2}
        \label{fig:subfig2}
    \end{subfigure}
    
    \vspace{0.3cm} 
    
    \begin{subfigure}[b]{1\columnwidth}
        \centering
        \includegraphics[width=\textwidth]{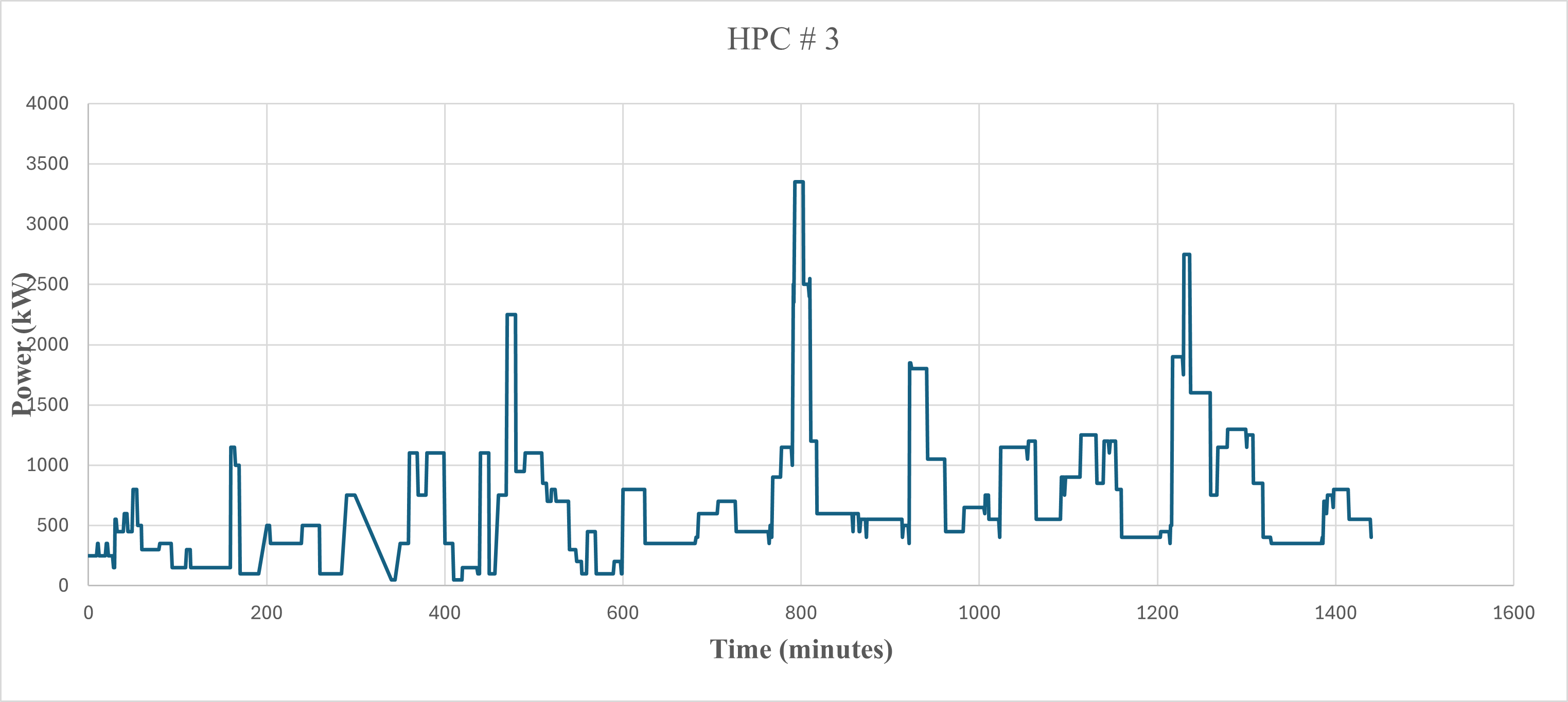}
        \caption{Load profile for HPCS 3}
        \label{fig:subfig3}
    \end{subfigure}
    
    \caption{Load profiles for High Power Charging Stations at three locations.}
    \label{fig:allfigures}
\end{figure}

HEVI-LOAD (Medium- and Heavy-Duty Electric Vehicle Infrastructure – Load Operation and Deployment) is a software tool developed by Lawrence Berkeley National Laboratory (US) that projects the charging infrastructure needs for MHDEVs, including the type, number, and location of charging stations. It provides visualizations and data to the county level, helping stakeholders plan for the electrification of commercial vehicle fleets. The Real-world data from HEVI-LOAD for 3 locations is used in this work to study the impact of HPCS on the secondary distribution system.\\
The raw data files contain the following columns: Time (minutes), Vehicle ID, Charging Power (kW), and Charging Time (minutes). There is an overlap of several vehicle IDs in the raw data file. This is not realistic because the number of charging ports at different charging stations is limited. Therefore, data is cleaned by applying the $M/M/c$ queue model, where $M$ is a Poisson arrival process with rate $\lambda$, $M$ is the Charging Time of the vehicle, and $c$ is the number of charging ports as shown in Fig.\ref{fig:mmc}. The vehicles are charged on a First Come First Served (FCFS) basis. The load profiles for 3 different locations in California are shown in Fig. \ref{fig:allfigures}

\subsection{IEEE 33-bus power distribution system}

\begin{figure}
    \centering
    \includegraphics[width=1\linewidth]{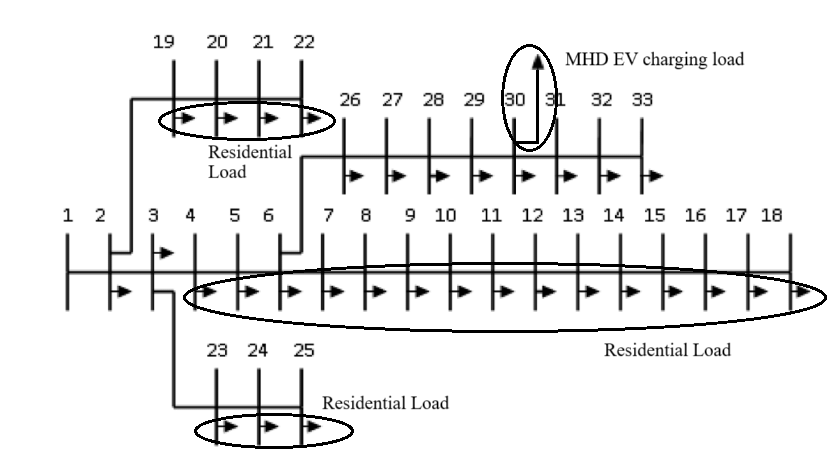}
    \caption{IEEE 33-bus distribution benchmark modified for case study}
    \label{fig:33bus}
\end{figure}

The IEEE 33-bus radial distribution system is a widely recognized benchmark for distribution system studies. This test system consists of 33 buses and 32 branches, operating at a nominal voltage of 12.66 kV. The benchmark system is modified to include time-varying residential loads obtained from \cite{b9} shown in Fig.\ref{fig:33bus}. The hourly residential load is interpolated to make the time granularity similar to the MHDEV load profile. AC Power flow simulations are run using Panda Power, an open-source Python library designed for power system modeling, analysis, and optimization. At each time step, Real Power $P_i$ and Reactive power $Q_i$ along with Voltage $V$ (p.u.) and angle $\theta$ is calculated as given in eq.\ref{eq:active_power} and eq.\ref{eq:reactive_power}.

\begin{equation}
P_i = V_i \sum_{j=1}^n V_j (G_{ij} \cos \theta_{ij} + B_{ij} \sin \theta_{ij})
\label{eq:active_power}
\end{equation}

\begin{equation}
Q_i = V_i \sum_{j=1}^n V_j (G_{ij} \sin \theta_{ij} - B_{ij} \cos \theta_{ij})
\label{eq:reactive_power}
\end{equation}
Where,
$P_i$ is the active power injected at bus $i$,
$Q_i$ is the reactive power injected at bus $i$,
$V_i$ is the voltage magnitude at bus $i$,
$V_j$ is the voltage magnitude at bus $j$,
$G_{ij}$ is the real part of the element in the bus admittance matrix corresponding to the i-th row and j-th column,
$B_{ij}$ is the imaginary part of the element in the bus admittance matrix corresponding to the i-th row and j-th column and 
$\theta_{ij}$ is the difference in voltage angle between bus $i$ and bus $j$ ($\theta_i - \theta_j$).

\section{Results and Discussion}
\begin{figure}[h]
    \centering
    \includegraphics[width=\columnwidth]{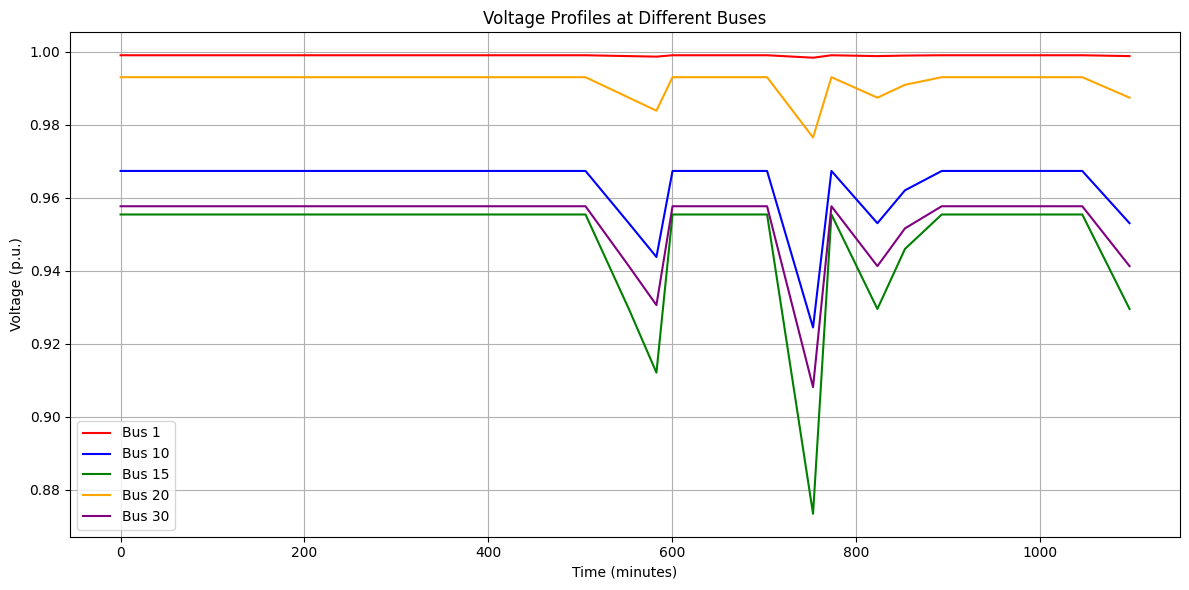} 
    \caption{Voltage (p.u.) values at different buses where MHDEV charging load is connected.}
    \label{fig:voltage_pu}
\end{figure}

Voltage stability indices are essential for assessing and maintaining the reliability of secondary distribution systems. The acceptable voltage range at distribution nodes is between 0.95 and 1.05 p.u. Figure \ref{fig:voltage_pu} illustrates the voltage values for the 24-hour time period. The buses were randomly selected to connect to the MHDEV charging load. The results show that MHDEV charging loads at nodes near the feeder, such as Bus 1 and Bus 10, demonstrate stable performance by maintaining voltage values within the acceptable range. However, buses located farther from the feeder, such as Bus 15, Bus 20, and Bus 30, violate the lower voltage threshold around the 570-minute mark that corresponds to 9:15 AM, when the EV charging load at Location 1 reaches 1.4 MW. This suggests that distribution systems with a radial topology require significant line upgrades to accommodate MHDEV charging loads effectively. Moreover, smart charge management and using co-located distributed energy resources (DERs) can also contribute to better voltage stability in the distribution grid. A future work in smart charge management will address the solution to the voltage instability.

\begin{table}[]
    \centering
\caption{Minimum voltage on different buses }
\label{tab:my_label}
    \begin{tabular}{|c|c|} 
    \hline 
    Bus Number  &  Minimum Voltage(p.u)\\
    \hline
    1 & 0.9984 \\ \hline 
    10 & 0.9245 \\ \hline 
    15 & 0.8734 \\ \hline 
    20 & 0.9766 \\ \hline 
    30 & 0.9081  
     \\ \hline\end{tabular} 
\end{table}

\section{Conclusion}
This work demonstrates that the integration of megawatt-scale charging stations for medium and heavy-duty electric vehicles (MHDEVs) poses significant challenges to the stability of secondary distribution systems. Our analysis, utilizing real-time data from the HEVI-LOAD software on a benchmark IEEE 33-bus distribution system, reveals substantial violations of per-unit voltage values at nodes farther from the feeder. This calls for upgrades to the distribution grid or utilization of co-located generation resources at the HPCS.

\section*{Acknowledgment}

This work is done in collaboration with Lawrence Berkeley National Lab, California, USA.

\vspace{12pt}

\end{document}